\documentclass[english,prb, manuscript]{revtex4}
\usepackage{times}
\usepackage[T1]{fontenc}
\usepackage[latin1]{inputenc}
\usepackage{verbatim}
\usepackage{amsmath}
\usepackage{color}
\usepackage{graphicx}

\makeatletter

\newcommand{\noun}[1]{\textsc{#1}}
\newenvironment{lyxgreyedout}{\textcolor[gray]{0.8}\bgroup}{\egroup}

\usepackage{babel}
\makeatother
\begin{document}

\newcommand{\sgrad}{\nabla_{S}}

\newcommand{\degree}{^{\circ}}

\newcommand{\dif}[1]{\vec{#1}-\vec{#1}'}

\newcommand{\dd}[2]{\delta^{#2}\left(#1-#1'\right)}

\newcommand{\vd}[2]{\delta^{#2}\left(\vec{#1}-\vec{#1'}\right)}

\newcommand{\ds}[1]{\delta\left(#1-#1'\right)}

\newcommand{\grad}{\vec{\nabla}}

\newcommand{\phik}[3]{e^{#3i#1\cdot#2}}

\newcommand{\intn}[4]{\int_{#1}^{#2}d^{n}\vec{#3}\left(#4\right)}

\newcommand{\height}{\mathcal{H}}

\newcommand{\h}{\height}

\renewcommand{\vec}{\mathbf}

\title{Anisotropy and Order of Epitaxial Self-Assembled Quantum Dots}

\author{Lawrence H. Friedman}

\affiliation{Dept. of Engineering Science and Mechanics, Pennsylvania State University,
212 Earth and Engineering Science Building, University Park, Pennsylvania
16802}

\date{\today}

\begin{abstract}
Epitaxial self-assembled quantum dots (SAQDs) represent an important
step in the advancement of semiconductor fabrication at the nanoscale
that will allow breakthroughs in electronics and optoelectronics.
In these applications, order is a key factor. Here, the role of crystal
anisotropy in promoting order during early stages of SAQD formation
is studied through a linear analysis of a commonly used surface evolution
model. Elastic anisotropy is used a specific example. It is found
that there are two relevant and predictable correlation lengths. One
of them is related to crystal anisotropy and is crucial for determining
SAQD order. Furthermore, if a wetting potential is included in the
model, it is found that SAQD order is enhanced when the deposited
film is allowed to evolve at heights near the critical surface height
for three-dimensional film growth.
\end{abstract}


\maketitle
Epitaxial self-assembled quantum dots (SAQDs) represent an important
step in the advancement of semiconductor fabrication at the nanoscale
that will allow breakthroughs in electronics and optoelectronics.
SAQDs are the result of linearly unstable film growth in strained
heteroepitaxial systems such as $\mbox{Si}_{x}$$\mbox{Ge}_{1-x}$/Si
and $\mbox{In}_{x}\mbox{Ga}_{1-x}\mbox{As}$/GaAs and other systems.
SAQDs have great potential for electronic and optoelectronic applications.
In these applications, order is a key factor. There are two types
of order, spatial and size. Spatial order refers to the regularity
of SAQD dot placement, and it is necessary for nano-circuitry applications.
Size order refers to the uniformity of SAQD size which determines
the voltage and/or energy level quantization of SAQDs. It has been
observed that crystal anisotropy can have a beneficial effect on SAQD
order.~\cite{Liu:2003kx} Here, the role of crystal anisotropy in
promoting order during early stages of SAQD formation is studied through
a linear analysis of a commonly used surface evolution model.~\cite{Spencer:1993vt,Liu:2003kx,Zhang:2003tg,Tekalign:2004jh,Friedman:2006bc}

The linear analysis addresses the initial stages of SAQD formation
when the nominally flat film surface becomes unstable and transitions
to three-dimensional growth. This early stage of SAQD growth determines
the initial seeds of order or disorder in an SAQD array, and can be
analyzed analytically. A dispersion relation as in~\cite{Spencer:1993vt,Golovin:2003ms}
is used, and two predictable correlation lengths that grow as the
square-root of time (Eqs.~\ref{eq:lcor} and~\ref{eq:ancor}) are
found. One of the correlation lengths results from crystal anisotropy.
This length plays a limiting role in the initial order of SAQD arrays,
and it is shown that anisotropy is crucial for creating a lattice-like
structure that is most technologically useful. This method of analysis
can be extended for use in any {}``nucleationless'' model of SAQD
growth, although here, the specific instance of elastic anisotropy
is treated as elastic anisotropy is the most easily estimated. Anisotropy
of surface energy may also effect SAQD order, and a similar analysis
would result. 

At later stages when surface fluctuations are large, non-linear dynamics
come into play. At this stage, there is a natural tendency of SAQDS
to either order or ripen.~\cite{Golovin:2003ms,Liu:2003kx} Ripening
systems will tend to have increased disorder as time progresses, while
ordering systems will tend to order only slightly due to critical
slowing down.~\cite{Wang:2004dd} Also, it is important to note that
while an ordering system might be {}``ordered'' when compared to
other non-linear phenomena such as convection roles, etc.~\cite{Cross:1993ti},
the requirements for technological application are much more stringent.
In a numerical investigation, it was found that SAQDs have enhanced
order due to crystal anisotropy, but soon become disordered from ripening.~\cite{Liu:2003kx}
In any case, an understanding of the order during the initial stages
of SAQD growth is essential to further investigation of the final
SAQD array order.

As in~\cite{Zhang:2003tg,Golovin:2003ms,Liu:2003kx,Spencer:1993vt},
a wetting energy is included in the analysis. The wetting potential
ensures that growth takes place in the Stranski-Krastanow mode: a
3D unstable growth occurs only after a critical layer thickness is
achieved, and a residual wetting layer persists. Although somewhat
controversial, the physical origins and consequences of the wetting
potential are discussed in~\cite{Beck:2004yq,Spencer:1993vt}. The
analysis presented here is quite general, and one can exclude or neglect
the effect of the wetting potential by simply setting it to zero.
That said, if the wetting potential is real, the present analysis
shows that it beneficial to SAQD order to grow near the critical layer
thickness.

The remainder of this report is organized as follows. First, stochastic
initial conditions are discussed. Second, the evolution of a single
mode for the isotropic case is discussed. Third, the resulting correlation
functions and correlation lengths are derived for the isotropic case.
Fourth, the analysis is repeated for the anisotropic case using elastic
anisotropy as an example. Finally, a representative numerical example
is presented using parameters appropriate for Ge dots grown on Si. 

To analyze resulting SAQD order, the mathematical model must include
stochastic effects. For simplicity, stochastic initial conditions
with deterministic time evolution are used. This method of analysis
yields a correlation function that is used to characterize SAQD order.

In this model, the film height $\h$ is a function of lateral position
$\vec{x}$ and time $t$. The film height is treated as an average
film height ($\bar{\h}$) with surface fluctuations $h(\vec{x},t)$
, \begin{equation}
\mathcal{H}=\bar{\mathcal{H}}+h(\vec{x},t).\label{eq:total_height}\end{equation}
In this way, $\bar{\h}$ functions as a control parameter\cite{Cross:1993ti}
physically signifying the amount of available material per unit area
to form SAQDs, and $h(\vec{x},t)$ evolves via surface diffusion giving
the the resulting surface profile. Order is then analyzed using using
the spatial correlation function, $\left\langle h(\vec{x},t)h(\vec{0},t)\right\rangle $
and the corresponding spectrum function $\left\langle h_{\vec{k}}(t)h_{\vec{k}'}(t)^{*}\right\rangle $. 

An initially flat surface is in unstable equilibrium, and it is necessary
to perturb it to produce SAQDs. Therefore, stochastic initial conditions
are implemented by letting $h(\vec{x},0)$ in Eq.~\ref{eq:total_height}
be random white noise. Specifically, $h(\vec{x},0)$ is assumed to
be sampled from a normal distribution such that\begin{equation}
\left\langle h(\vec{x},0)\right\rangle =0\mbox{, and }\left\langle h(\vec{x},0)h(\vec{x}',0)\right\rangle =\Delta^{2}\delta^{d}\left(\vec{x}-\vec{x}'\right)\label{eq:xmean}\end{equation}
where $\Delta$ is the noise amplitude of dimension {[}length]$^{1+d/2}$,
$d$ is the dimension of the surface, and $\delta^{d}(\vec{x})$ is
the $d\mbox{-dimensional}$ Dirac Delta function. Much of the following
analysis uses the Fourier transform with the convention, $h_{\vec{k}}(t)=(2\pi)^{-d}\int d^{d}\vec{k}\,\exp(-i\vec{k}\cdot\vec{x})h(\vec{x},t).$
The mean and two-point correlation functions for $h_{\vec{k}}(0)$
are \begin{equation}
\left\langle h_{\vec{k}}(0)\right\rangle =0\mbox{, and }\left\langle h_{\vec{k}}(0)h_{\vec{k}'}(0)^{*}\right\rangle =\frac{\Delta^{2}}{(2\pi)^{d}}\delta^{d}\left(\vec{k}-\vec{k}'\right).\label{eq:kmean}\end{equation}
%
{}

The deterministic evolution of a single Fourier component is determined
by surface diffusion with a diffusion potential $\mu(\vec{x},t)$.\cite{Zhang:2003tg,Spencer:1993vt}
This model is phenomenological in nature, but contains the essential
elements of SAQD formation. Thus, it is an adequate, but not overly
complex starting point for the investigation of the effects of crystal
anisotropy. Furthermore, models of this nature can be derived from
atomic scale simulations.~\cite{Haselwandter:2006hy} At any instant
in time, the growing film is described by the curve $\mathcal{H}(\vec{x},t)$.
Using Eq.~\ref{eq:total_height} to decompose the film height, \begin{eqnarray*}
\partial_{t}h(\vec{x},t) & = & \vec{\nabla}\cdot(\mathcal{D}\vec{\nabla}\mu(\vec{x},t;\bar{\h}));\\
d\bar{\h}/dt & = & Q,\end{eqnarray*}
 where $\mu$ depends on $\bar{\h}$, and $Q$ is the flux of new
material onto the surface. 

%
{}

The appropriate diffusion potential $\mu$ must produce Stranski-Krastanow
growth. Thus, it must incorporate the elastic strain energy density
$\omega$ that destabilizes a planar surface, the surface energy density
$\gamma$ that stabilizes planar growth and a wetting potential $W(\mathcal{H})$
that ensures substrate wetting. The %
{} simplest form that gives the appropriate behavior is

\begin{equation}
\mu=\Omega\left(\omega-\kappa\gamma+n_{z}W'(\h)\right)\label{eq:cp2}\end{equation}
similar to~\cite{Liu:2003kx,Golovin:2003ms,Tekalign:2004jh,Friedman:2006bc}
where $\Omega$ is the atomic volume, $\kappa$ is the total surface
curvature, and $n_{z}$ is the vertical component of the unit surface
normal $\hat{n}$. The strain energy density $\omega$ is found using
isotropic linear plane strain elasticity. In general, $\omega$ is
a function of $x$, and it is a non-local functional of the entire
surface profile $\mathcal{H}(x)$.%
{}

Consider first, the the one-dimensional and two-dimensional isotropic
cases. Following~\cite{Golovin:2003ms,Spencer:1993vt,Tekalign:2004jh},
the surface diffusion potential, (Eq.~\ref{eq:cp2}) is expanded
to first order in the film height fluctuation $h$. The elastic energy
$\omega$ is calculated using linear isotropic elasticity. It is a
non-local function of $h(\vec{x})$; thus, it is useful to work with
the Fourier transform. In the isotropic case, the linearized diffusion
potential (Eq.~\ref{eq:chem_pot}) depends only on the wave vector
magnitude $k=\left\Vert \vec{k}\right\Vert $ . Thus, \begin{eqnarray}
\mu_{\vec{k}} & = & f(k,\bar{\height})h_{\vec{k}},\mbox{ with}\label{eq:chem_pot}\\
f(k,\bar{\height}) & = & \Omega\left(-2M(1+\nu)\epsilon_{0}^{2}k+\gamma k^{2}+W''(\bar{\h})\right).\label{eq:fiso}\end{eqnarray}
%
{}%
{}%
{}%
{}%
{}In the anisotropic case, there will also be a dependence on the wave
vector direction $\theta_{\vec{k}}$.

%
{}The time dependence of the film height has a simple solution if there
is no additional flux of material ($Q=0$, and $\bar{\h}$ is constant).\begin{eqnarray}
h_{\vec{k}}(t) & = & h_{\vec{k}}(0)e^{\sigma_{k}t}\mbox{, with}\label{eq:simple_time}\\
\sigma_{k} & = & -\mathcal{D}k^{2}f(k,\bar{\h}),\label{eq:dispersion}\end{eqnarray}
where $\sigma_{k}$ is the \emph{dispersion relation} and depends
only on the wavevector magnitude $k$. %
{} Modes with positive $\sigma_{k}$ are grow unstably, while modes
with negative values of $\sigma_{k}$ decay.

The important features of $\sigma_{k}$ are most easily recognized
using a characteristic wavenumber is $k_{c}=2M(1+\nu)\epsilon_{0}^{2}/\gamma$
and a characteristic time $t_{c}=(\mathcal{D}\Omega\gamma k_{c}^{4})^{-1}$.~\cite{Spencer:1993vt,Golovin:2003ms}

\begin{equation}
\sigma_{k}=t_{c}^{-1}\alpha^{2}(\alpha-\alpha^{2}-\beta).\label{eq:dp}\end{equation}
where the shorthand $\alpha=k/k_{c}$ and $\beta=W''(\bar{\h})/(\gamma k_{c}^{2})$
is used.

The dispersion relation (Eq.~\ref{eq:dp}) has a peak at $k_{0}=\alpha_{0}k_{c}$
where \[
\alpha_{0}=\frac{1}{8}\left(3+\sqrt{9-32\beta}\right).\]
Expanding $\sigma_{k}$ about $k_{0}$, \begin{eqnarray}
\sigma_{k}\approx\sigma_{0}-\frac{1}{2}\sigma_{2}\left(k-k_{0}\right)^{2}\mbox{ where }\nonumber \\
\sigma_{0}=\frac{1}{4t_{c}}\alpha_{0}^{2}\left(\alpha_{0}-2\beta\right),\quad\sigma_{2}=k_{c}^{-2}t_{c}^{-1}(3\alpha_{0}-4\beta).\label{eq:G2}\end{eqnarray}
Thus,\begin{equation}
h_{\vec{k}}(t)\approx h_{\vec{k}}(0)e^{\sigma_{0}t-\frac{1}{2}\sigma_{2}t\left(k-k_{0}\right)^{2}}.\label{eq:TEapprox}\end{equation}
%
{} 

Now, the statistical correlation functions and correlation lengths
that characterize order are derived. Using Eqs.~\ref{eq:simple_time}
and~\ref{eq:TEapprox} along with the stochastic initial conditions
(Eqs.~\ref{eq:xmean} and~\ref{eq:kmean}) , the mean value of $h_{\vec{k}}(t)$
is \[
\left\langle h_{\vec{k}}(t)\right\rangle =\left\langle h_{\vec{k}}(0)\right\rangle e^{\sigma_{k}t}=0,\]
so that the mean surface perturbation is simply $0$ for all time
and all $\vec{k}$. However, the mean-square surface perturbations
as characterize by the second-order correlation function~\cite{Zwanzig:2001zf}
can be large,\begin{eqnarray}
\left\langle h_{\vec{k}}(t)h_{\vec{k}'}(t)^{*}\right\rangle  & = & \left\langle h_{\vec{k}}(0)h_{\vec{k}'}(0)^{*}\right\rangle e^{(\sigma_{k}+\sigma_{k'})t}\nonumber \\
 & = & \frac{\Delta^{2}}{(2\pi)^{d}}\vd{k}{d}e^{2\sigma_{k}t},\label{eq:k-cor}\end{eqnarray}
using Eq.~\ref{eq:kmean}. The real space correlation function can
be found by taking the inverse Fourier transform of Eq.~\ref{eq:k-cor},

\begin{eqnarray}
\left\langle h(\vec{x},t)h(\vec{x}',t)^{*}\right\rangle  & = & \int d^{d}\vec{k}\int d^{d}\vec{k}'\, e^{i\vec{k}\cdot\vec{x}-i\vec{k}'\cdot\vec{x}'}\left\langle h_{\vec{k}}(t)h_{\vec{k}'}(t)^{*}\right\rangle \nonumber \\
 & = & \frac{\Delta^{2}}{(2\pi)^{d}}\int d^{d}\vec{k}\, e^{i\vec{k}\cdot\left(\vec{x}-\vec{x}'\right)}e^{2\sigma_{k}t}\label{eq:hcor}\end{eqnarray}
where integration over $\vec{k}'$ is simple due to the $\vd{k}{d}$
in Eq.~\ref{eq:k-cor}. 

Using Eq.~\ref{eq:G2}, \[
e^{2\sigma_{k}t}\approx e^{2\sigma_{0}t-\frac{1}{2}(2\sigma_{2}t)(k-k_{0}^{2})}\]
which is peaked at $k=k_{0}$. This form suggests that the real-space
correlation function is periodic with a gaussian envelope that has
a standard deviation of\begin{equation}
L_{cor}=\sqrt{2\sigma_{2}t}\label{eq:lcor}\end{equation}
$L_{cor}$ is the \emph{correlation length} of the film-height profile,
and characterizes the degree of order of the SAQD array. For example,
in one dimension ($d=1)$,\begin{eqnarray*}
 &  & \left\langle h(x,t)h(0,t)^{*}\right\rangle \dots\\
 &  & \dots=\frac{\Delta^{2}}{2\pi}\sum_{\pm}\int_{0}^{\infty}dk\, e^{2\sigma_{0}t-\frac{1}{2}L_{cor}^{2}(k-k_{0})^{2}\pm ikx}\\
 &  & \dots\approx\frac{2\Delta^{2}}{(2\pi L_{cor}^{2})^{1/2}}e^{2\sigma_{0}t-\frac{1}{2}(x/L_{cor})^{2}}\cos(k_{0}x)\end{eqnarray*}
an approximation that is valid if $k_{c}L_{cor}\gg1$. Thus, $L_{cor}$
gives the length scale over which dots will be ordered, and this scale
grows as $t^{1/2}$. As $t\rightarrow\infty$, \[
\left\langle h(x,t)h(0,t)^{*}\right\rangle =\frac{2\Delta^{2}}{(2\pi L_{cor}^{2})^{1/2}}e^{2\sigma_{0}t}\cos(k_{0}x),\]
and the entire array should be perfectly ordered.

\begin{figure}
\includegraphics[width=1.75in]{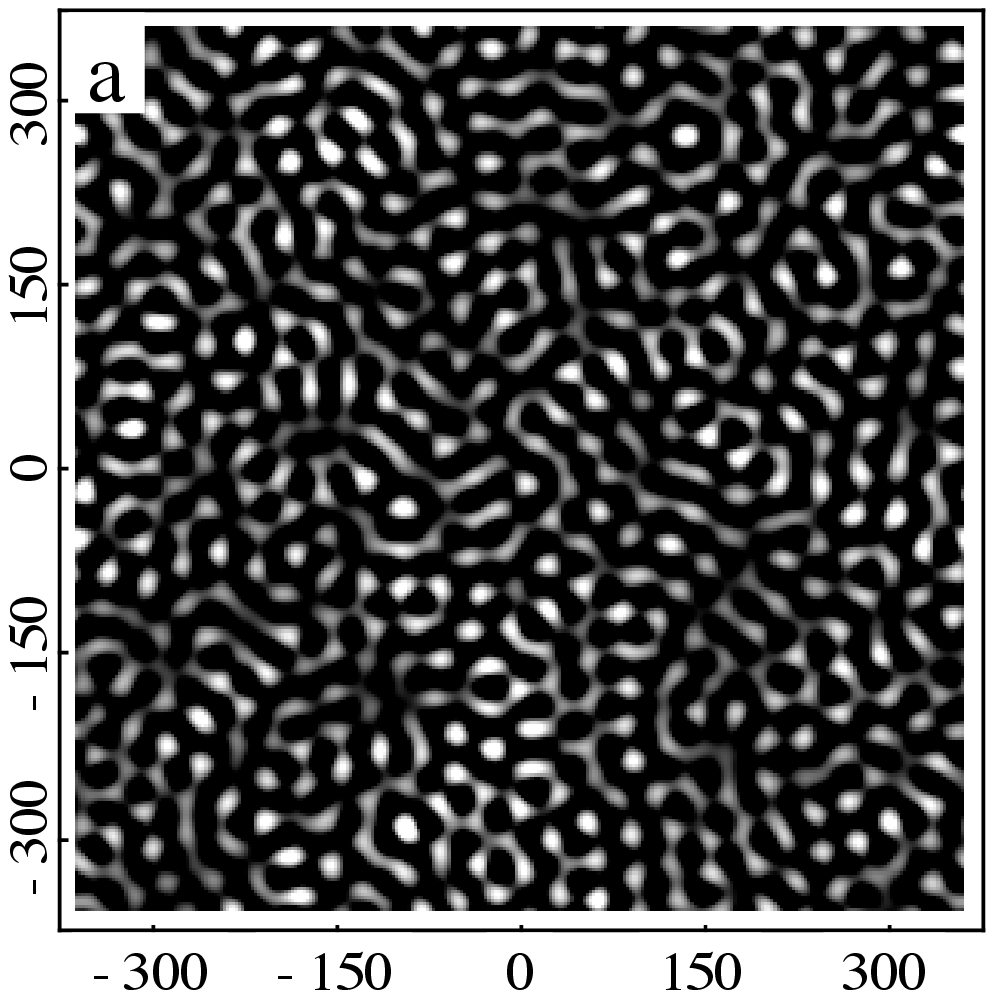}\includegraphics[width=1.75in]{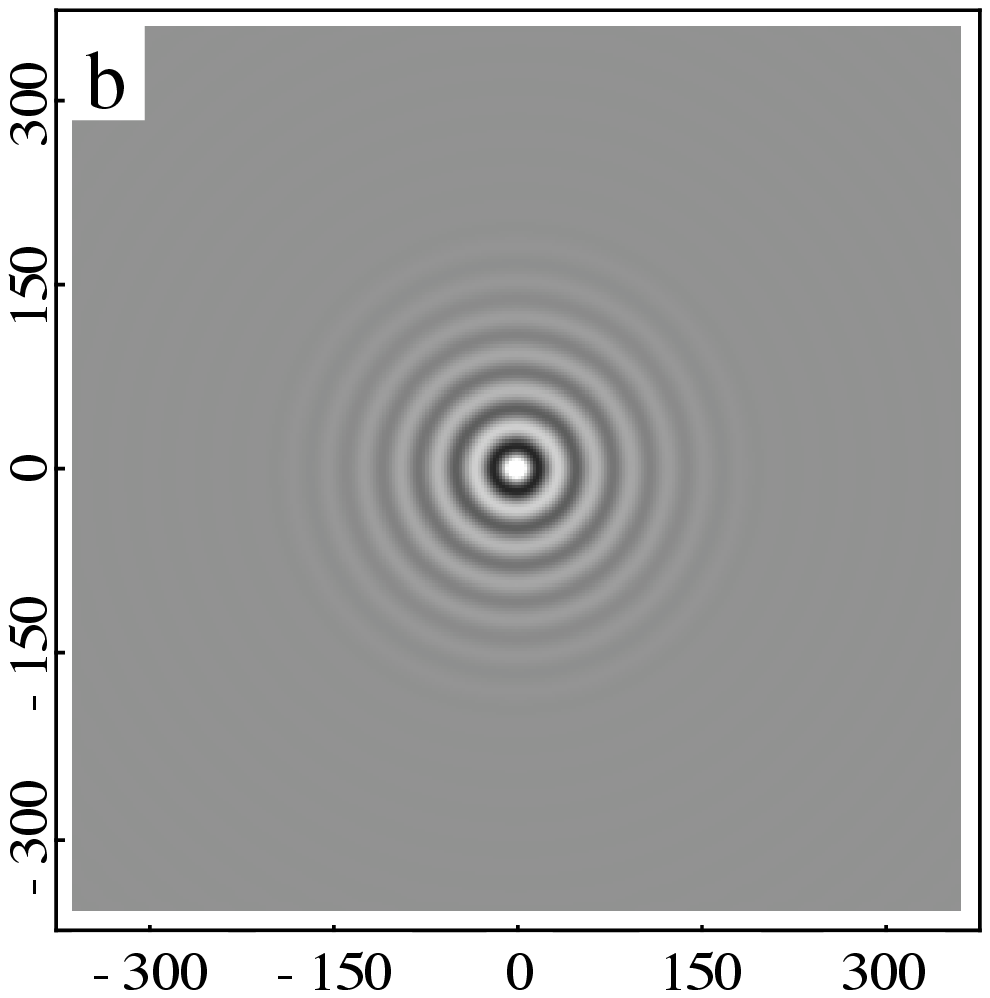}

\includegraphics[width=1.75in]{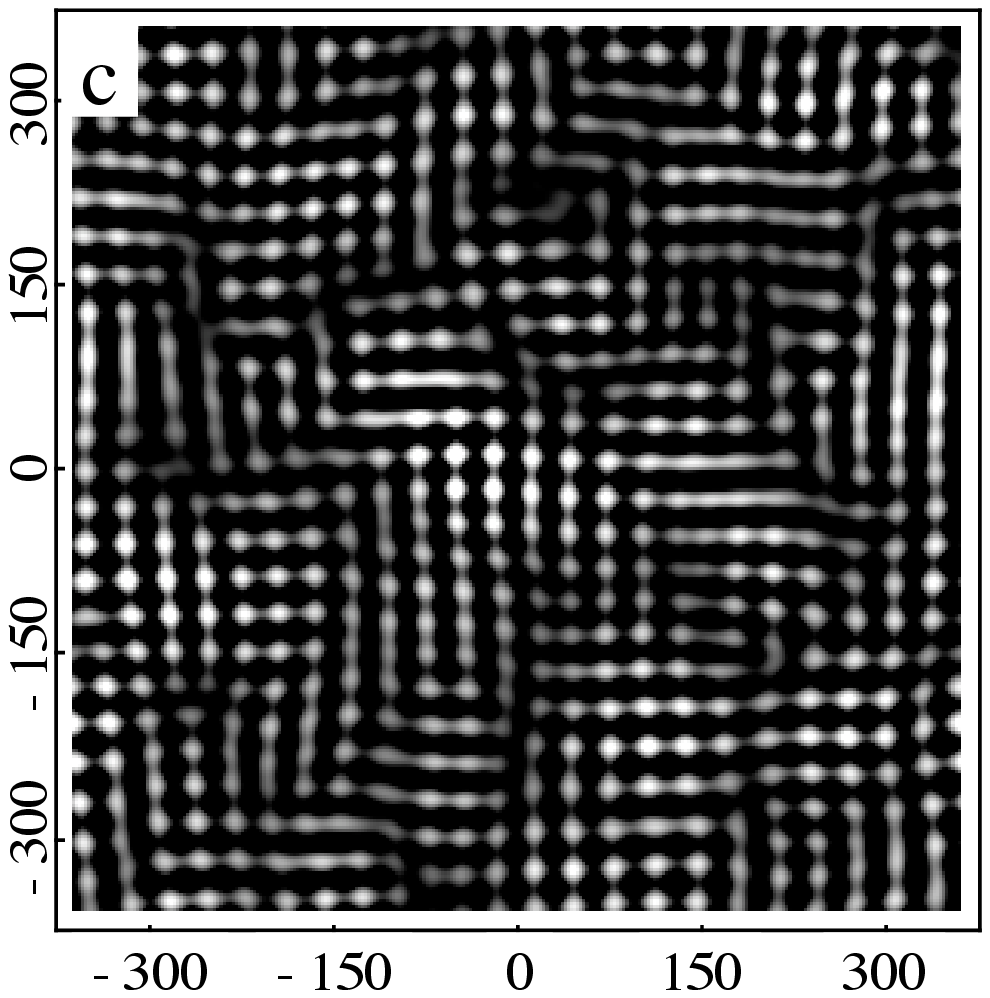}\includegraphics[width=1.75in]{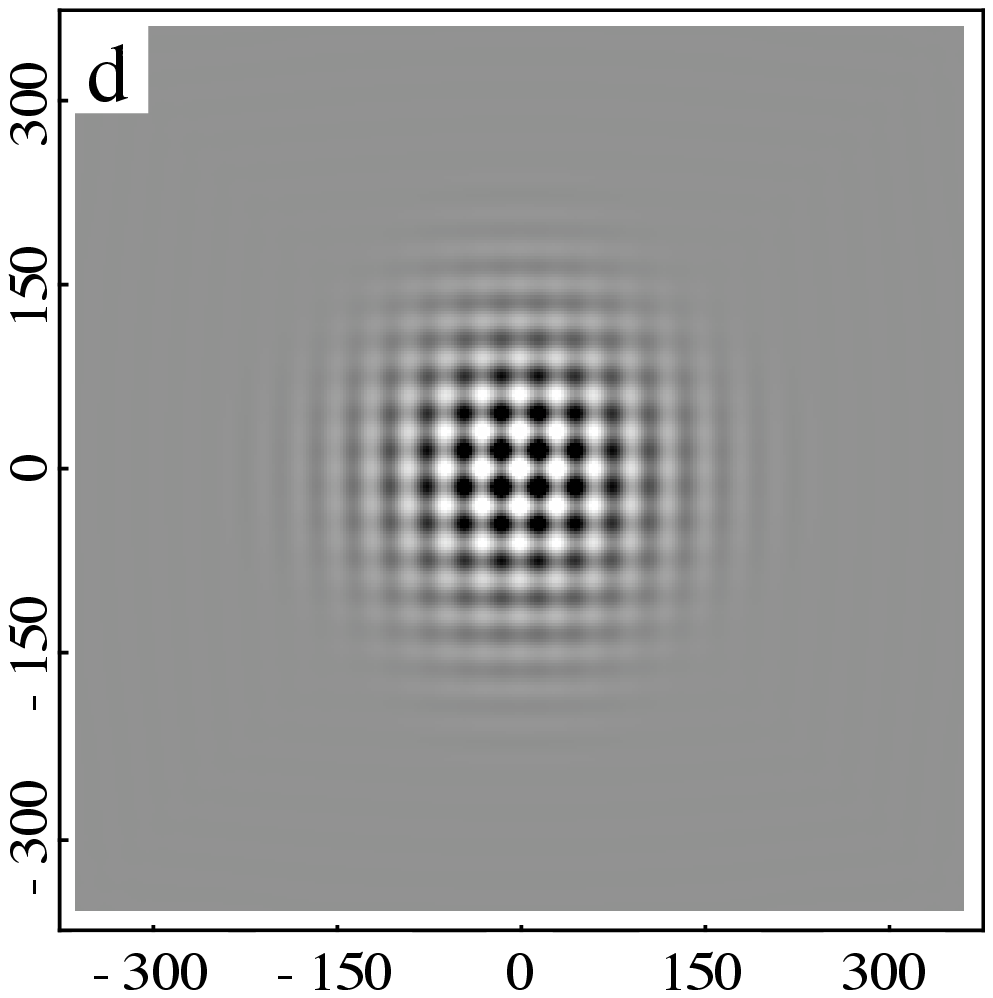}

\caption{\emph{\label{cap:DensityPlots}Density plots of surface profiles
and correlation functions at} $t/t_{c}=471.$ The scale is in nm,
and the physical parameters are listed in the text. Plots (a) and
(b) are a sample surface profile and correlation function respectively
for an elastically isotropic material. Plots (c) and (d) are a surface
profile and correlation function for an elastically anisotropic material
with 4-fold symmetry. Surface profile plots saturate black at a minimum
of $\h=\bar{\h}$ for clarity. White dots indicate surface peaks.
Correlation function plots span $\pm$ the maximum value / 2.}
\end{figure}

The situation in two-dimensions, however, is less friendly. As $t\rightarrow\infty$,
$L_{cor}\rightarrow\infty$, and $e^{-\frac{1}{2}L_{cor}^{2}(k-k_{0})}\approx(2\pi/L_{cor}^{2})^{1/2}\delta(k-k_{0})$
\begin{eqnarray*}
 &  & \left\langle h(\vec{x},t)h(\vec{0},t)^{*}\right\rangle \dots\\
 &  & \dots=\frac{\Delta^{2}}{(2\pi)^{3/2}L_{cor}}\int d^{2}\vec{k}\, e^{2\sigma_{0}t+i\vec{k}\cdot\vec{x}}\delta(k-k_{0})\\
 &  & \dots=\frac{\Delta^{2}k_{0}}{(2\pi L_{cor}^{2})^{1/2}}e^{2\sigma_{0}t}J_{0}(k_{0}\parallel\vec{x}\parallel)\end{eqnarray*}
Thus, the two-dimensional isotropic case has statistical order at
large times, but does not yield SAQD lattices as does the one-dimensional
case (see Fig.~\ref{cap:DensityPlots}a and~b).

Now, consider the effects of crystal anisotropy. For example, let
the elastic energy term $\omega$ have $N$-fold symmetry while the
other terms are assumed isotropic. Then, the growth rate depends on
the both the wave vector magnitude $k$ (and thus on $\alpha=k/k_{c}$)
and direction $\theta_{\vec{k}}$ so that, $\sigma_{k}\rightarrow\sigma_{\vec{k}}$.
Naturally, the anisotropic elastic energy term in the growth rate
$\sigma_{\vec{k}}$ depends on the specific anisotropic elastic constants,
but the general qualitative effect of elastic anisotropy on SAQD growth
kinetics can be investigated without incorporating a detailed elastic
calculation at this time. Thus, the present work provides motivation
for more detailed calculation. A reasonable way to estimate how the
elastic energy term varies with direction ($\theta_{\vec{k}}$) is
to assume a low order harmonic form with the proper rotational symmetry.
The simplest such guess is\begin{eqnarray*}
\sigma_{\vec{k}} & = & t_{c}^{-1}\alpha^{2}\left(\alpha\left(1-\epsilon\sin^{2}\left(N\theta_{\vec{k}}/2\right)\right)-\alpha^{2}-\beta\right),\end{eqnarray*}
where $\epsilon$ parameterizes the importance of the directional-dependence.
More precise calculations using anisotropic elastic constants of real
materials (similar to ref.~\cite{Obayashi:1998fk}) will be given
in future work.

$\sigma_{\vec{k}}$ has peaks at $N$ wave vectors, \[
\vec{k}_{n}=k_{0}\left(\left(\cos\theta_{n}\right)\hat{e}_{x}+\left(\sin\theta_{n}\right)\hat{e}_{y}\right)\]
with $\theta_{n}=2\pi(n-1)/N$. Around each peak, $\vec{k}$ can be
decomposed in the direction parallel ($k_{\parallel})$ and perpendicular
($k_{\perp}$) to $\vec{k}_{n}$. Using this decomposition and expanding
$e^{2\sigma_{\vec{k}}t}$ about each peak,\begin{eqnarray}
 &  & e^{2\sigma_{\vec{k}}t}\approx\sum_{n=1}^{N}\exp\left(2\sigma_{n}t\right);\label{eq:expansion}\\
 &  & 2\sigma_{n}t=2\sigma_{0}t-\frac{1}{2}L_{\parallel}^{2}(k_{\parallel}-\alpha_{0}k_{c})^{2}-\frac{1}{2}L_{\perp}^{2}k_{\perp}^{2};\label{eq:tauGn}\\
 &  & L_{\parallel}=\sqrt{2\sigma_{2}t},\mbox{ and }L_{\perp}=\sqrt{(N^{2}\alpha_{0}\epsilon t_{c}^{-1}k_{c}^{-2})t}.\label{eq:ancor}\end{eqnarray}
Note that $L_{\parallel}$ is the same as $L_{cor}$ for the isotropic
case. Eq.~\ref{eq:expansion} is valid if $k_{c}L_{\parallel}\gg1,\mbox{ and }k_{c}L_{\perp}\gg1$.
Using, Eq.~\ref{eq:hcor}, but noting that $\sigma_{\vec{k}}$ now
depends on both the magnitude and direction of $\vec{k}$, along with
Eqs.~\ref{eq:expansion}, and~\ref{eq:tauGn},

\begin{eqnarray}
 &  & \left\langle h(\vec{x},t)h(\vec{0},t)^{*}\right\rangle =\frac{\Delta^{2}}{(2\pi)^{2}}\frac{2\pi}{L_{\parallel}L_{\perp}}\dots\nonumber \\
 &  & \dots\sum_{n=1}^{N/2}e^{2\sigma_{0}t-\frac{1}{2}\left(L_{\parallel}^{-2}x_{\parallel}^{2}+L_{\perp}^{-2}x_{\perp}^{2}\right)}2\cos\left(k_{0}x_{\parallel}\right),\label{eq:hcorA}\end{eqnarray}
where $x_{\parallel}=(\cos\theta_{n})x+(\sin\theta_{n})y$, and $x_{\perp}=(-\sin\theta_{n})x+(\cos\theta_{n})y$.
Thus, the same tendency to long-range order as for the one-dimensional
case is present (see Fig.~\ref{cap:DensityPlots}c and~d).

As a numerical example, consider Ge grown on Si. Both the isotropic
approximation and an estimated elastically anisotropic case (with
$N=4$, and $\epsilon=0.1$) are treated. Neglecting the difference
in elastic properties of the Si substrate, $E_{\text{Ge}}=1.361\times10^{12}\mbox{ dyne/cm}^{2}$,
$\nu_{\text{Ge}}=0.198$, $\epsilon_{0}=-0.0418$, $\Omega=2.27\times10^{-23}\mbox{cm}^{3}$,
$\gamma=1927\mbox{ erg/cm}^{2}$, and $W(\h)=4.75\times10^{-6}/\h\mbox{ erg/cm}^{3}$.
The resulting biaxial modulus is $M=1.697\times10^{12}\text{ dyne/cm}^{2}$,
characteristic wavenumber is $k_{c}=0.369\text{ nm}^{-1}$ and critical
film height is $\bar{\h}_{c}=1.132\text{ nm}\approx4\mbox{ ML}$.
If the film is grown to a thickness of $\bar{\h}=\h_{c}+0.25\mbox{ ML}\approx1.203\text{ nm}$,
and then allowed to evolve, $\beta=0.2078$, $\alpha_{0}=0.5664$,
$k_{0}=0.209\text{ nm}^{-1}$, $\sigma_{0}=0.01206t_{c}^{-1}$, $\sigma_{2}=0.867k_{c}^{-2}t_{c}^{-1}$,
$L_{\parallel}=0.746k_{0}^{-1}(t/t_{c})^{1/2}$, and $L_{\perp}=0.539k_{0}^{-1}(t/t_{c})^{1/2}$.
Note that the unspecified diffusivity $\mathcal{D}$ has been absorbed
into $t_{c}$. The film will stay in the linear regime as long as
the surface fluctuations stay small. For this purpose, let {}``small''
mean $1\mbox{ ML}=2.83\times10^{-8}\mbox{cm}$. Once the fluctuations
become {}``large'', individual dots will begin to form, and a nonlinear
analysis becomes necessary. It is useful to know the correlation lengths
at this time.

To find the correlation lengths, one must choose the initial height
fluctuation intensity $\Delta$ and then calculate the time for fluctuations
to become {}``large''. The initial fluctuation intensity is somewhat
arbitrary, but $\Delta=8.02\times10^{-16}\mbox{ cm}^{2}$ gives an
average fluctuation of $1\mbox{ ML}$ over a patch $1\mbox{ ML}^{2}$
and seems appropriate. Next Eqs.~\ref{eq:ancor} and~\ref{eq:hcorA}
are used to find $t$ for which the r.m.s. fluctuations become large
$h_{rms}=\left\langle |h(0,t)|^{2}\right\rangle ^{1/2}=a_{0}$. There
are two solutions, $t/t_{c}=5.53\times10^{-3}$, and $t/t_{c}=471$.
The first solution is an artifact of the white noise initial conditions
and occurs during an initial shrinking of the surface height variance;
thus, the second solution is taken. At $t/t_{c}=471$, the correlation
lengths are found using Eq.~\ref{eq:ancor}, $L_{cor}=L_{\parallel}=77.5\mbox{ nm}$,
and $L_{\perp}=56.0\mbox{ nm}$. The smaller correlation length gives
$k_{0}L_{\perp}/\pi=3.73$, so a patch of about 4 dots across is expected
to be reasonably well ordered.

A numerical simulation of linear size $l=24(2\pi/k_{0})=722\text{ nm}$
can be easily performed. The discrete initial conditions $h_{\vec{k}}$(0)
are taken from a normal distribution of zero mean and variance $\left\langle h_{\vec{k}}(0)h_{\vec{k}'}^{*}(0)\right\rangle _{discrete}=(\Delta^{2}/l^{2})\delta_{\vec{k}\vec{k}'}$,
where $\delta_{\vec{k}\vec{k}'}$ is the Kronecker delta, and each
vector component of $\vec{k}$ takes discrete values $2\pi m/l$ with
$m$ an integer. These components then evolve via Eq.~\ref{eq:simple_time}.
The results of an isotropic and anisotropic simulation along with
the corresponding real-space correlation functions are shown in Fig.~\ref{cap:DensityPlots}.
These plots clearly demonstrate the importance of anisotropy in producing
long range order.

Figs.~and~\ref{cap:DensityPlots}c and~d appear to agree qualitatively
with observations of nucleationless growth of Ge$_{x}$Si$_{1-x}$
nanostructures on Si~\cite{Berbezier:2003mw,Brunner:2002gf,Gao:1999ve},
although mostly with $x<0.5$. Typical  observed dot arrangements
appear to correspond to lower values of $\beta$ than 0.2078 used
for the example as they are quasiperiodic but less ordered than Fig.~\ref{cap:DensityPlots}c.
Quantitative reporting of correlation lengths would assist comparison
and possibly enable better tuning of phenomenological models to experiments.
InP/InGaP nanostructures reported in~\cite{Bortoleto:2003zh} appear
similar.

From the analysis of the isotropic model, it is clear that long range
statistical order (long correlation lengths) require tight distributions
in reciprocal space. This long range statistical order is achieved
in the large time limit, but this statistical order is insufficient
to produce a well-ordered array of SAQDs. This lack of usable order
is reflected in the real-space two-point correlation function of the
isotropic model. However, in the one-dimensional and two-dimensional
anisotropic cases, there is tendency to form a lattice after a long
time. In the anisotropic case, there are two correlation lengths that
characterize observed SAQD array order. Analytic formulas for these
correlation lengths have been given for a model with elastic simplified
anisotropy, but the general conclusions and method of analysis should
apply to any source of anisotropy. Additionally, from a simple form
of the wetting potential it is observed that at film heights just
above the critical threshold $\h_{c}$, the correlation lengths can
grow quickly while the height fluctuations grow slowly; thus order
is enhanced. Availability of measured SAQD correlation lengths would
help to improve this analysis and to engineer more ordered quantum
dot arrays.


\begin{thebibliography}{10}

\bibitem{Liu:2003kx}
P.~Liu, Y.~W. Zhang, C.~Lu.
\newblock {\em Phys. Rev. B}, 67:165414, 2003.

\bibitem{Spencer:1993vt}
B.~J. Spencer, P.~W. Voorhees, S.~H. Davis.
\newblock {\em J. Appl. Phys.}, 73(10):4955--4970, 1993.

\bibitem{Zhang:2003tg}
Y.W. Zhang, A.F. Bower, P.~Liu.
\newblock {\em Thin solid films}, 424:9--14, 2003.

\bibitem{Tekalign:2004jh}
W.~T. Tekalign,  B.~J. Spencer.
\newblock {\em J. Appl. Phys.}, 96(10):5505--5512, 2004.

\bibitem{Friedman:2006bc}
Lawrence~H. Friedman,  Jian Xu.
\newblock {\em Appl. Phys. Lett.}, 88:093105, 2006.

\bibitem{Golovin:2003ms}
A.~A. Golovin, S.~H. Davis, P.~W. Voorhees.
\newblock {\em Phys. Rev. E}, 68:056203, 2003.

\bibitem{Wang:2004dd}
Yu~U. Wang, Yongmei~M. Jin, Armen~G. Khachaturyan.
\newblock {\em Acta Mater.}, 52:81--92, 2004.

\bibitem{Cross:1993ti}
M.~C. Cross,  P.~C. Hohenberg.
\newblock {\em Rev. Mod. Phys.}, 65(3):851--1112, 1993.

\bibitem{Beck:2004yq}
M.~J. Beck, A.~van~de Walle, M.~Asta.
\newblock {\em Phys. Rev. B}, 70(20):205337, Nov 2004.

\bibitem{Haselwandter:2006hy}
C.~Haselwandter,  D~VVedensky.
\newblock {\em 2006 APS March Meeting}, 2006.

\bibitem{Zwanzig:2001zf}
Robert Zwanzig.
\newblock {\em Nonequilbrium Statistical Mechanics}.
\newblock Oxford University Press, New York, 2001.

\bibitem{Obayashi:1998fk}
Y.~Obayashi,  K.~Shintani.
\newblock {\em J. Appl. Phys.}, 84(6):3141, 1998.

\bibitem{Berbezier:2003mw}
I.~Berbezier, A.~Ronda, F.~Volpi, A.~Portavoce.
\newblock {\em Surf. Sci.}, 531:231--243, 2003.

\bibitem{Brunner:2002gf}
Karl Brunner.
\newblock {\em Rep. Prog. Phys.}, 65(1):27--72, 2002.

\bibitem{Gao:1999ve}
H.~J. Gao,  W.~D. Nix.
\newblock {\em Annu. Rev. Mater. Sci.}, 29:173--209, 1999.

\bibitem{Bortoleto:2003zh}
J.~R.~R. Bortoleto, H.~R. Gutierrez, M.~A. Cotta, J.~Bettini, L.~P. Cardoso,
  M.~M.~G. de~Carvalho.
\newblock {\em Appl. Phys. Lett.}, 82(20):3523--3525, 2003.

\end{thebibliography}
\end{document}